\documentclass[fleqn,usenatbib]{mnras}

\usepackage{newtxtext,newtxmath}

\usepackage[T1]{fontenc}
\DeclareRobustCommand{\VAN}[3]{#2}
\let\VANthebibliography\thebibliography
\def\thebibliography{\DeclareRobustCommand{\VAN}[3]{##3}\VANthebibliography}

\usepackage{graphicx}	
\usepackage{amsmath}	

\usepackage{xcolor}

\usepackage{url}
\usepackage{lipsum} 
\usepackage[bottom]{footmisc} 
\usepackage{tabu}
\graphicspath{{./}{figures/}}
\usepackage{appendix}
\usepackage[utf8]{inputenc}
\usepackage{graphicx}
\usepackage{fancyhdr}
\usepackage[flushleft]{threeparttable}

\title[DNN for GRB redshift]{Deep Neural Networks for Estimation of Gamma-Ray Burst Redshifts}

\author[Aldowma \& Razzaque]{
Tamador Aldowma,$^{1,2}$\thanks{E-mail: tamtam2030@gmail.com}
Soebur Razzaque,$^{1,3,4}$\thanks{E-mail: srazzaque@uj.ac.za}
\\
$^{1}$Centre for Astro-Particle Physics (CAPP) and Department of Physics, University of Johannesburg, PO Box 524, Auckland Park 2006, South Africa\\
$^{2}$Department of Astronomy and Meteorology, Faculty of Science and Technology, Omdurman Islamic University, PO Box 382, Omdurman, 14415, Sudan\\
$^{3}$Department of Physics, The George Washington University, Washington, DC 20052, USA \\
$^{4}$National Institute for Theoretical and Computational Sciences (NITheCS), South Africa 
}

\date{Accepted XXX. Received YYY; in original form ZZZ}
\pubyear{2015}

\begin{document}
\label{firstpage}
\pagerange{\pageref{firstpage}--\pageref{lastpage}}
\maketitle

\begin{abstract}
While the available set of Gamma-ray Burst (GRB) data with known redshift is currently limited, a much larger set of GRB data without redshift is available from different instruments. This data includes well-measured prompt gamma-ray flux and spectral information. We estimate the redshift of a selection of these GRBs detected by {\it Fermi}-GBM and Konus-{\it Wind} using Machine Learning techniques that are based on spectral parameters. We find that Deep Neural Networks with Random Forest models employing non-linear relations among input parameters can reasonably reproduce the pseudo-redshift distribution of GRBs, mimicking the distribution of GRBs with spectroscopic redshift. Furthermore, we find that the pseudo-redshift samples of GRBs satisfy (i) Amati relation between the peak photon energy of the time-averaged energy spectrum in the cosmological rest frame of the GRB ${E}_{\rm i, p}$ and the isotropic-equivalent radiated energy ${E}_{\rm iso}$ during the prompt phase; and (ii) Yonetoku relation between ${E}_{\rm i, p}$ and isotropic-equivalent luminosity ${L}_{\rm iso}$, both measured during the peak flux interval.         
\end{abstract}

\begin{keywords}
(stars:) gamma-ray burst: general - methods: data analysis
\end{keywords}



\section{Introduction}\label{sec:section1}
The discovery of a Gamma-ray Burst (GRB) dates back to 1967 when the Vela military satellite first detected it \citep{1973ApJ...182L..85K}. Since then, numerous space telescopes have observed GRBs, while ground-based instruments have monitored the afterglow emission. These observations have revealed GRBs to be the most powerful explosions in electromagnetic wavebands ever discovered, with the majority of energy being emitted in gamma rays in the keV to MeV range \citep{1995PASP..107.1145F}. GRBs can be categorized into two main types based on their spectral hardness and characteristic duration ${T}_{90}$ \citep{1993ApJ...413L.101K}: long GRBs (LGRBs) with ${T}_{90} > 2$~s, typically associated with the explosive demise of massive stars, and short GRBs (SGRBs) with ${T}_{90} < 2$~s, associated with mergers of binary neutron stars or a black hole and a neutron star \citep[for reviews see, e.g.,][]{1999PhR...314..575P}. The parameter ${T}_{90}$ represents the time within which 90\%  \citep{2012ApJS..199...18P} of the collected gamma-ray flux is accumulated. SGRBs generally exhibit lower fluence on average but similar peak flux compared to LGRBs. 

By analyzing optical afterglows and identifying host galaxies, redshifts of GRBs have been measured up to $z = 9.4$ \citep{Cucchiara2011_736}. This provides compelling evidence that GRBs have a cosmological origin and represent the most powerful explosions since the Big Bang \citep{2000astro.ph..4176F}. The sample of GRBs with measured redshift, however, is still limited and depends on following up afterglows using optical telescopes on the ground with spectroscopic capabilities. Statistical analysis techniques based solely on GRB data can potentially be used to estimate a pseudo redshift and thus can increase the count of GRBs with redshift \citep{2003A&A...407L...1A}. Over the past two decades, researchers have discovered several empirical relationships between observable parameters of prompt emission.  One such example is the Amati correlation, denoted as $E_{\rm iso}/{10}^{52} {\rm erg} = k ({E}_{\rm i, p}/{\rm keV})^m$ \citep{Amati2002_81A, Amati2006_372}. This correlation relates the intrinsic peak energy ${E}_{\rm i, p}$, at which the prompt gamma-ray spectral energy distribution $\nu{f}_{\nu}$ peaks in the cosmological rest frame, to the isotropic-equivalent gamma-ray energy release ${E}_{\rm iso}$ in the cosmological source frame. Additionally, Yonetoku relation describes the relationship between ${E}_{\rm i, p}$ and the isotropic-equivalent luminosity ${L}_{\rm iso}$ during the peak flux interval  \citep{2004ApJ...609..935Y, 2014Ap&SS.351..267Z, 2019NatCo..10.1504I, 2020ApJ...900...33P}. These correlations, if tight, can be used to calculate the pseudo-redshift of a GRB and possibly use them as cosmological standard candles \citep[see, eg.,] []{2005ApJ...627....1F, 2006NJPh....8..123G, 2008MNRAS.391..577A, Basilakos2008_391, 2011MNRAS.415.3580D, Wang2016_585, 2016ApJ...831L...8G, Demianski2017_693, Dirirsa2019, Khadka2021, Demianski2021, Kumar2023}. Estimating pseudo-redshift using these phenomenological relations, however, requires calculating luminosity distance $d_L$ using a particular cosmological model. There is, therefore, a circularity in logic in case one uses these phenomenological relations to also make GRBs cosmological standard candles \citep[see, eg.,][]{Li2007, Dirirsa2019, 2019MNRAS.486L..46A}. Studies have also been carried out on selection effects in these relations \cite[see, eg.,][]{Ghirlanda2008_387, Butler2009, Nava2009_1133, Ghirlanda2012_422}.  

Recently, Machine Learning (ML) approaches have unveiled intriguing correlations among observable properties of GRBs with available redshift and to estimate GRB redshift using observed data \citep[see, eg.,][]{2016MNRAS.458.3821U, Dainotti2019, 2020grbg.conf..141D, 2022JCAP...04..016E, 2021MNRAS.503.4581L}. \citet{2022JCAP...04..016E}, e.g., used Recurrent Neural Network (RNN) and a
Bayesian Neural Network (BNN) to calibrate observed properties of high redshift GRBs. \citet{2020grbg.conf..141D}, e.g., used SuperLearner \citep{vanderLaanPolleyHubbard+2007} to calculate pseudo redshift of a sample of {\it Swift} GRBs based on their observed prompt and afterglow characteristics. This kind of study does not use a cosmology model to calculate pseudo-redshift, thus avoiding the circularity problems in GRB phenomenological relations, such as the Amati, Yonetoku, Ghirlanda, etc., which may also be used to estimate redshift but require calculation of the luminosity distance using a cosmology model to derive parameters of those relations. With a reliable and large sample of GRB pseudo-redshift inferred from ML techniques, it might be possible to use GRBs as cosmological standard candles.

In this study, we have employed regression techniques in Deep Neural Networks  (DNN) and Random Forest algorithms to calculate the pseudo-redshift of a large sample of LGRBs detected by {\it Fermi}-GBM \citep{VonKienlin2020}. We have trained and tested our ML  models on the spectral properties of prompt emission for LGRBs detected by {\it Fermi}-GBM \citep{VonKienlin2020, 2021ApJ...913...60P} and by Konus-{\it Wind} (KW) \citep{2021ApJ...908...83T} with measured redshift. The best-fit models are then used to infer the pseudo-redshift of LGRBs based on their spectral properties. We have performed statistical tests on the pseudo-redshift samples and explored if they satisfy the Amati and Yonetoku correlations.  This paper is organized as follows. In Sec.\ \ref{sec:data} we define our data samples and analysis methods using ML techniques. In Sec.\ \ref{sec:results} we present results on pseudo-redshift samples and perform various tests. We discuss results in Sec.\ \ref{sec:discussion} and present conclusions in Sec.\ \ref{sec:conclusions}.

\section{Data Selection and Analyses}\label{sec:data}
In our analysis, we employ various observed features of GRBs to make predictions about their redshift. We utilize data sets obtained from the Fermi-GBM catalog\footnote{\url{https://heasarc.gsfc.nasa.gov/W3Browse/fermi/fermigbrst.html}} \citep{VonKienlin2020} which covered periods from 2008 to 2018. In a span of 10 years, Fermi-GBM triggered on a total of 2,356 GRBs \citep{VonKienlin2020}. Detailed information regarding the spectral properties and energy of these GRBs can be found in \citep{2021ApJ...913...60P}. Among the detected GRBs, GBM has identified 135 with known redshift. In our analysis, we exclude 13 SGRBs, as they are not included in our current study. Therefore, the GBM catalog \citep{2021ApJ...913...60P} includes a total of 122 LGRBs. Additionally, we have incorporated five GRBs detected by GBM, namely GRB~080913B; GRB~110721A; GRB~150120A; GRB~180703A, and GRB~180720B, in our input data. 

GRB~090423A \citep{2009Natur.461.1254T, 2009Natur.461.1258S} stands out as the farthest LGRB in our sample at a redshift of ${z} = 8.2$. The Fermi-GRB database contains more than 3,000 events, with approximately 16\% of them classified as SGRBs. Using the {\tt gbm} tool\footnote{\url{https://fermi.gsfc.nasa.gov/ssc/data/analysis/gbm/}} \citep{GbmDataTools}, we obtained all GBM data without redshift from the Fermi-GBM burst catalog. We have also downloaded data from the KW catalog\footnote{\url{http://www.ioffe.ru/LEA/zGRBs/part2/index.html}} \citep{2021ApJ...908...83T}, which encompasses the time period from 2005 to 2018.  In our sample selection, we have included a subset of 127 LGRBs from GBM and 338 LGRBs from KW, all of which have measured redshift information. Furthermore, the energy ranges for spectral fitting fall within the Fermi-GBM detection range of 8-1000 keV, and for KW, it ranges from 80-1200 keV. 

To perform our analysis, we utilize spectral fitting parameters obtained from the Band model \citep{1993ApJ...413..281B} and Comptonized model \citep{2009PASP..121.1279S} for GRB spectra. These parameters correspond to various spectral characteristics, such as the peak bolometric flux ${P}_{\rm bolo}$ for calculating the isotropic peak luminosity ${L}_{\rm iso}$ and the bolometric fluence ${S}_{\rm bolo}$ for calculating the isotropic energy ${E}_{\rm iso}$. We use two sets of parameters, namely the ``peak flux'' and ``Fluence'' parameters as shown in Table ~\ref{tab:table one}, and there are a few percentage differences between the peak flux and Fluence parameters in the Fermi-GBM and KW data, which we have ignored. Each set includes spectral parameters, peak flux or fluence, and burst duration $T_{90}$. These choices are motivated by the phenomenological Amati \citep{Amati2002_81A} and Yonetoku \citep{2004ApJ...609..935Y} correlations, although we do not use $L_{\rm iso}$ or $E_{\rm iso}$ that require using the redshift and a cosmological model. Similarly, we also do not use the intrinsic peak energy of the spectral fits that are used in the Amati and Yonetoku correlations. These choices are aimed to avoid biases in our analysis as we do not know {\it a priori} what relations might exist among different input parameters.     

Below, we describe two commonly employed models for fitting the prompt emission spectra of GRBs.

\paragraph*{Band Model:} 
This model comprises of two power laws characterized by spectral indices $\alpha$ and $\beta$ connected with an exponential function, along with the spectral peak energy ${E}_{p}$ measured in keV. The Band model offers a flexible and versatile approach to describing the complex spectral shapes observed in GRBs. The power laws capture the behavior of the spectrum at low and high energies, while the spectral peak energy represents the energy at which the emission is most intense. The model is given by
\begin{equation}
\resizebox{\linewidth}{!}{$
{N}_{\text{Band}}(E) = {A}_{\text{Band}} \left\{\begin{array}{l}
\left(\frac{E}{100 \text{ keV}}\right)^\alpha \exp\left[- \frac{E(2+\alpha)}{{E}_{p}}\right], \\ \text{if } E \leq {E}_{b} ; \\
\\
\left(\frac{E}{100 \text{ keV}}\right)^\beta \exp(\beta - \alpha)\left[- \frac{{E}_{p}}{100 \text{ keV}} \frac{\alpha - \beta}{2 + \alpha}\right]^{\alpha - \beta}, \\ \text{if } E > {E}_{b}
\end{array}
\right.$}
\label{eq:Band}
\end{equation}
where ${A}_{\rm Band}$ is the amplitude, and ${E}_{b} \equiv {E}_{p}(\alpha - \beta)/(2 + \alpha)$. These parameters, together with the bolometric peak flux $P_{\rm bolo}$ and fluence $S_{\rm bolo}$ calculated using the Band model are input parameters in our analyses and are listed in Table ~\ref{tab:table one}.

\paragraph*{Comptonized Model:} The Comptonized (Comp) model is a power-law with an exponential cutoff and is given by 

\begin{eqnarray} 
{N}_{\rm Comp} = {A}_{\rm Comp} \left(\frac{E}{100 ~ {\rm keV}}\right)^\alpha \exp\left[- (2 + \alpha) \frac{E}{{E}_{p}}\right] \,.
\label{eq:Comp}
\end{eqnarray}
Here ${A}_{\rm Comp}$ is the amplitude, $\alpha$ is the photon index and ${E}_{p}$ is the peak energy. Again, these are input parameters in our analyses together with $P_{\rm bolo}$ and $S_{\rm bolo}$ calculated using the Comp model, and are also listed in Table~\ref{tab:table one}. 

Afterward, we proceed to exclude GRBs with reported errors on spectral parameters exceeding 100\% and those that lacked ${E}_{p}$ values in all spectral models. We have also excluded GRBs best-fitted with the Band model for which ${\beta} \ge -2$, indicating the absence of a peak. This exclusion criteria applies to both GBM and KW data, regardless of whether their redshift is known or not. We refer to Table \ref{tab:example_table_0} for more details on different samples used in our analysis. Note that, as the GBM and KW data sets share the same spectral information features obtained from the Comp and Band model fitting, it is possible to combine the two data sets for each model. However, the number of GRBs differs for each spectral model. In Section \ref{sec:results} and the corresponding figures there, we present detailed information about the number of bursts included in each different analysis.

\begin{table}
	\centering
	\caption{Input parameters for DNN models from the GBM and KW catalogs. Apart from the spectral parameters, we also use $T_{90}$, fluence and flux, computed for each spectral fit, in our analysis}
	\label{tab:table one}
	\begin{tabular}{lccr} 
  \hline
 \multicolumn{3}{c}{\bf Spectral Parameters} \\
 \hline
		\hline
		Bolometric & Peak Flux & Fluence \\
		Band Model& $\alpha, \beta, {E}_{p}, {P}_{\rm bolo}$ & $\alpha, \beta, {E}_{p}, {S}_{\rm bolo}$\\
		Comptonized Model & ${\alpha}, {E}_{p}, {P}_{\rm bolo}$ & ${\alpha}, {E}_{p}, {S}_{\rm bolo}$ \\
		\hline
	\end{tabular}
\end{table}

\begin{table}
    \caption{Number of GRBs with known redshift (indicated with $z$) and without redshift in different samples based on the spectral model, after excluding those GRBs with large errors on the spectral parameters.}
	\centering
	\label{tab:example_table_0}
	\begin{tabular}{lccr} 
  \hline
		Model &  Fermi-GBM & Konus-Wind  \\
		\hline
  \hline
		Band fluence - $z$ & 94 &  56 \\
        Band fluence   & 854 &  - \\
        Band flux - $z$ &  100 &  33\\
        Band flux & 647  &  -\\
        Band fluence and flux - $z$ &  94 & 89 \\
        Band fluence and flux & 647  & - \\
        \hline
        Comp fluence - $z$ & 116  &  166 \\
        Comp fluence &  1858 &   -\\
        Comp flux - $z$& 116  &  110 \\
        Comp flux & 1708  &  - \\
        Comp fluence and flux -$z$ & 116   &   276 \\
        Comp fluence and flux &  1708  &   - \\
        \hline
	\end{tabular}
\end{table}

\subsection{Application of Deep Neural Networks to GRB data}
\label{sec:maths} 
Neural Networks (NNs) are computational structures composed of interconnected nodes, commonly employed in supervised learning tasks. DNNs are a specific type of NN that consists of multiple layers capable of performing intricate computations \citep{2022arXiv220411786C}. 
In our analysis, we employed a deep learning algorithm using TensorFlow\footnote{\url{https://www.tensorflow.org/} \\ {\it Is an open-source machine learning library developed by Google.}} \citep{geron2022hands}, a popular framework for implementing and training DNNs. TensorFlow provides a flexible and efficient platform for constructing and optimizing NNs for regression tasks. The regression approach in deep learning involves training the DNN on a labeled data set, where the inputs are the observed features of GRBs, and the outputs are the corresponding redshift values. By feeding the training data through the DNN, the network adjusts its internal parameters to minimize the discrepancy between predicted redshifts and the true redshift values.

We have explored other ML techniques such as the Linear Regression and Gaussian Process Regression, which however resulted in overfitting of data.
The advantage of using DNNs for regression tasks is their capability to capture complex and non-linear relationships between input features and the target variable, in this case, the redshift. The multiple layers in DNNs allow for hierarchical representations and the learning of intricate patterns in the data. We have described various elements of the DNNs used in our analysis in Appendix~\ref{app:DNN}. We describe below data inputs for DNNs and the Stacking Ensemble method to obtain the final ML model.

\subsubsection{Input data}
The input data for our analysis consists of spectral and other parameters mentioned in Table~\ref{tab:table one}. These parameters include various characteristics of the GRB spectra, such as the peak flux, fluence, and spectral fitting parameters from the Band and Comptonized models. We have applied a logarithmic transformation to the variables ${T}_{90}$, ${E}_{p}$, ${S}_{\rm bolo}$, and ${P}_{\rm bolo}$. The target variable we aim to predict is the redshift of the GRB. To train our DNN models, each spectral model is separately used on the data sets. Additionally, we employ a combination of the two spectral models.

\subsubsection{Scaling Data for DNN Regressor}
In the context of using a DNN regressor, scaling the data is an important preprocessing step that helps to normalize the input features and ensure that they are on a similar scale \citep{2020arXiv200912836H}. Scaling the data is particularly crucial when working with features that have different ranges or units of measurement
The most commonly used scaling technique is known as ``Standardization'' or ``Z-score normalization.'' It involves transforming the data in such a way that the mean of the feature becomes zero and the standard deviation becomes one. The equation for standardization is as follows:
\begin{eqnarray}
    {X}_{\rm std} = \frac{X - \mu}{\sigma}
\end{eqnarray}
where ${X}_{\rm std}$ represents the standardized value of the feature, ${X}$ is the original value of the feature, ${\mu}$ the mean of the feature, and ${\sigma}$ the standard deviation of the feature. This process ensures that the features have zero mean and unit variance, allowing the DNN regressor to learn more effectively from the data.

\subsubsection{Stacking Ensemble}
\begin{figure*}
\begin{center}
\includegraphics[width=16cm]{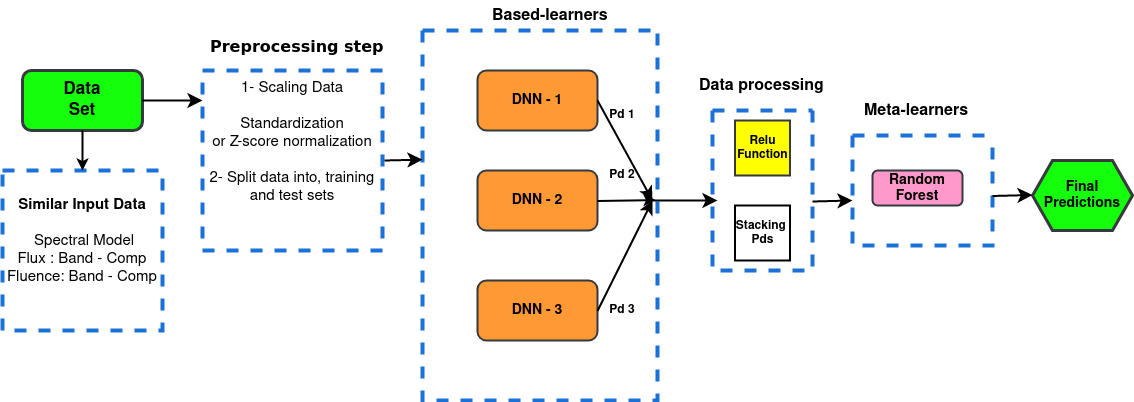}
\caption{Architecture of the Stacking Ensemble used in this work to find the best ML model for predicting GRB redshift. The predictions from the DNN models (Pd 1, Pd 2, and Pd 3) are stacked into one dataset using {\it numpy.stack}, and the final prediction is made with the meta-model \textit{Random Forest}.}
\label{fig:fig_A}
\end{center}  
\end{figure*}

To address the issue of overfitting and to improve the performance of our deep learning model, we employed a stacking ensemble technique \citep{2018PLoSO..1305872M, 2019arXiv190712659W, MOON2020109921}. 
The stacking model involves evaluating multiple individual models and combining their predictions by averaging the best results.
The architecture of a stacking ensemble used in our analyses is shown in Figure~\ref{fig:fig_A}. The ensemble consists of multiple DNN regressor models, called the Base-learners, which were trained with identical input data. Each of these models contributed its prediction to the ensemble, and the final ML models, called the Meta-learners, are Random Forests\footnote{\url{https://scikit-learn.org/stable/modules/generated/sklearn.ensemble.RandomForestRegressor.html}} from the \texttt{Scikit-learn} \citep{2011JMLR...12.2825P}. To ensure consistency, we used identical inputs and training procedures for each individual DNN model within the stacking ensemble. 
The commonly used metric to determine the performance of a Random Forest, which is the final ML model for predicting GRB redshift, is the coefficient of determination, also known as r-squared ($R^2$). Additionally, we have used another metric, the mean absolute error (MAE), that indicates the level of overfitting. In general closer the $R^2$ to one and closer the MAE to zero are indicators of better models.\footnote{\url{https://scikit-learn.org/stable/modules/generated/sklearn.ensemble.StackingRegressor.html}}

%

\section{Results}\label{sec:results}
\begin{figure*}
	\begin{center}
		\includegraphics[width=10.5 cm]{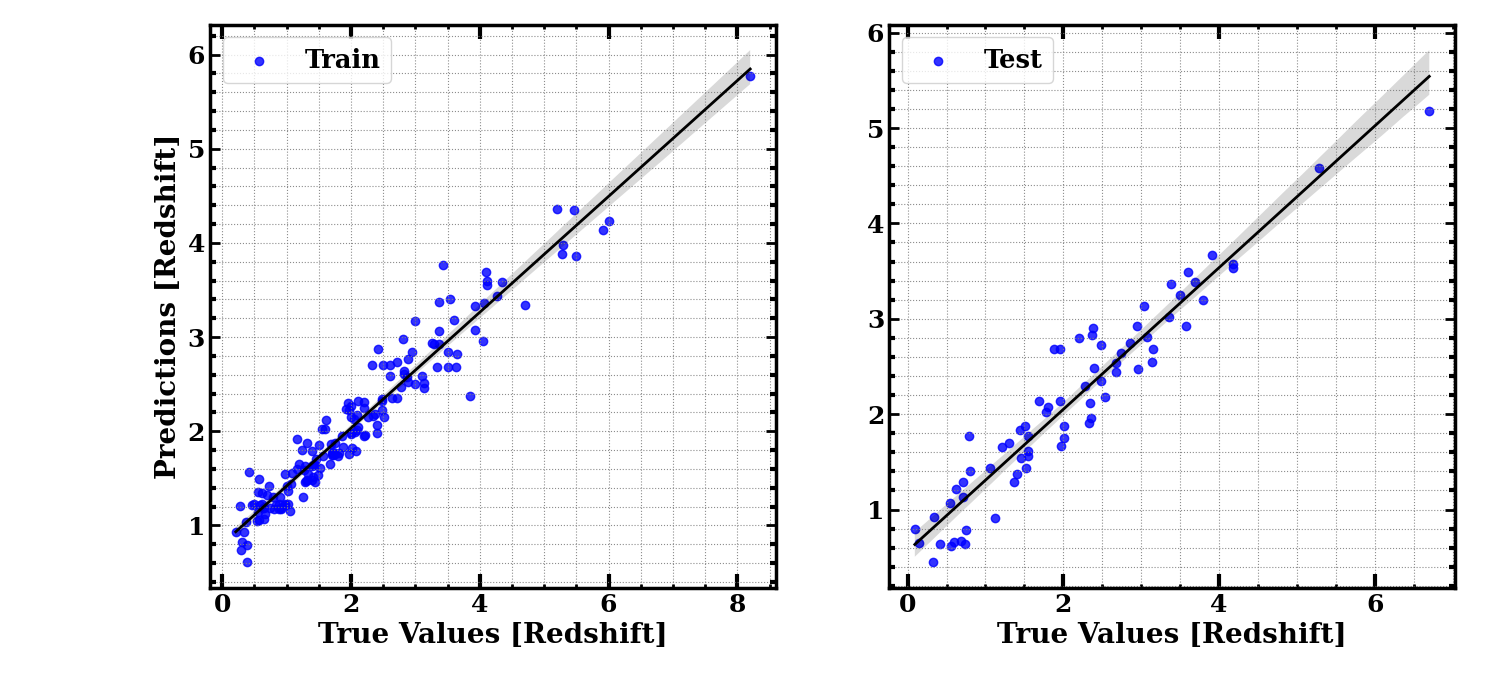}
        \includegraphics[width=10.5 cm]{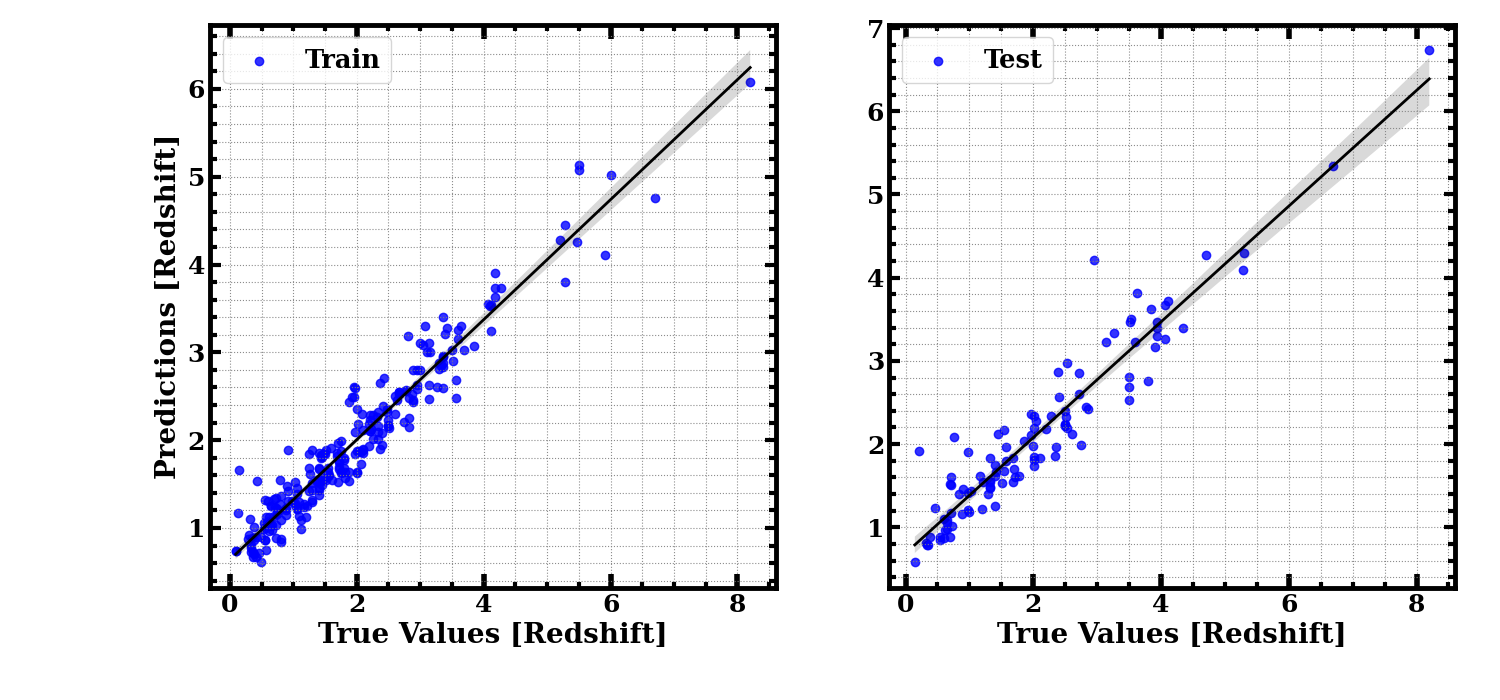}
        \includegraphics[width=10.5 cm]{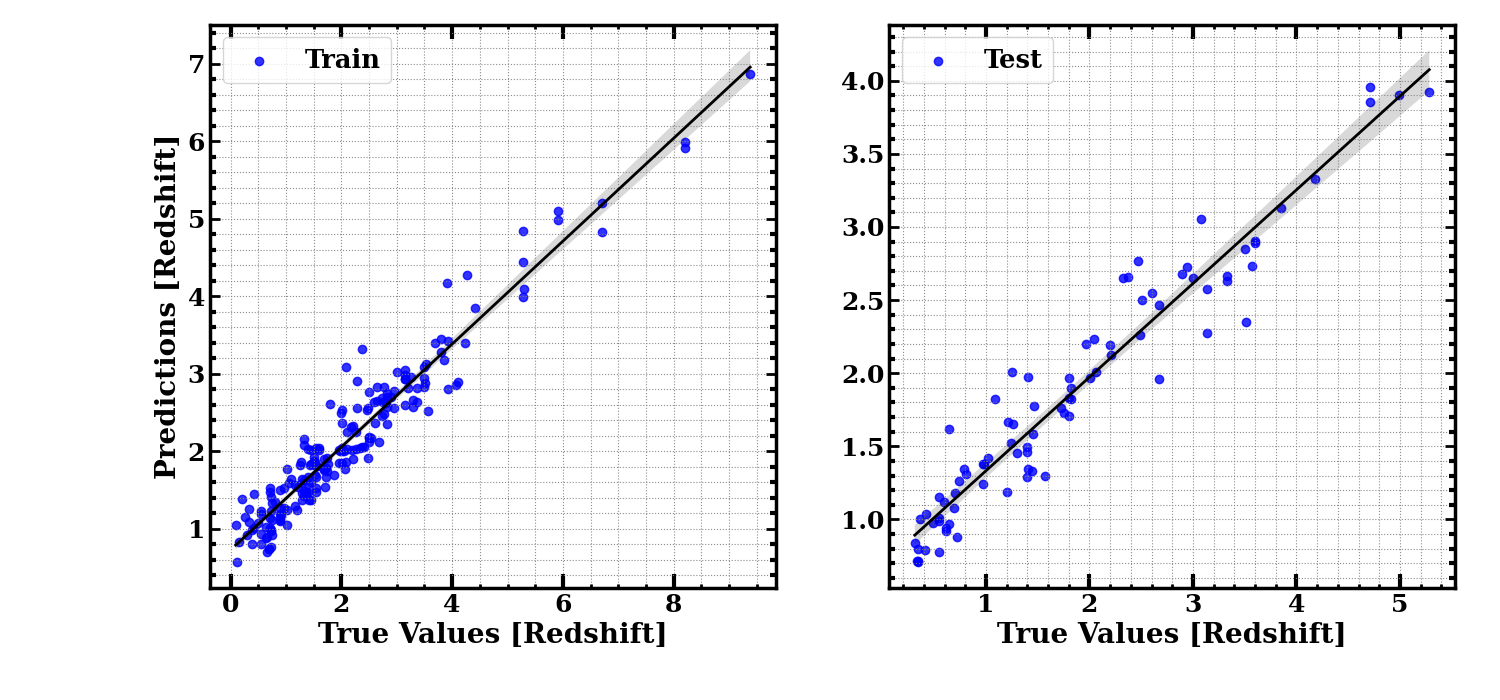}
		\caption{Predicted redshift from the {\it ensemble models} (DNNs + Random Forest)  vs.\ true redshift for train and test samples from the combined KW-GBM data with known redshift. Here we have used the ``Comp flux'' (top panels), ``Comp fluence and flux'' (middle panels), and ``Band fluence and flux'' (bottom panels) models mentioned in Table~\ref{tab:example_table}. The data sets used for the analyses are described in Table~\ref{tab:example_table_0}. The regression lines are shown with 95\% CL.}
		\label{fig:my_figure}
	\end{center}
\end{figure*}

\begin{figure*}
\begin{center}
\includegraphics[width=10.5cm]{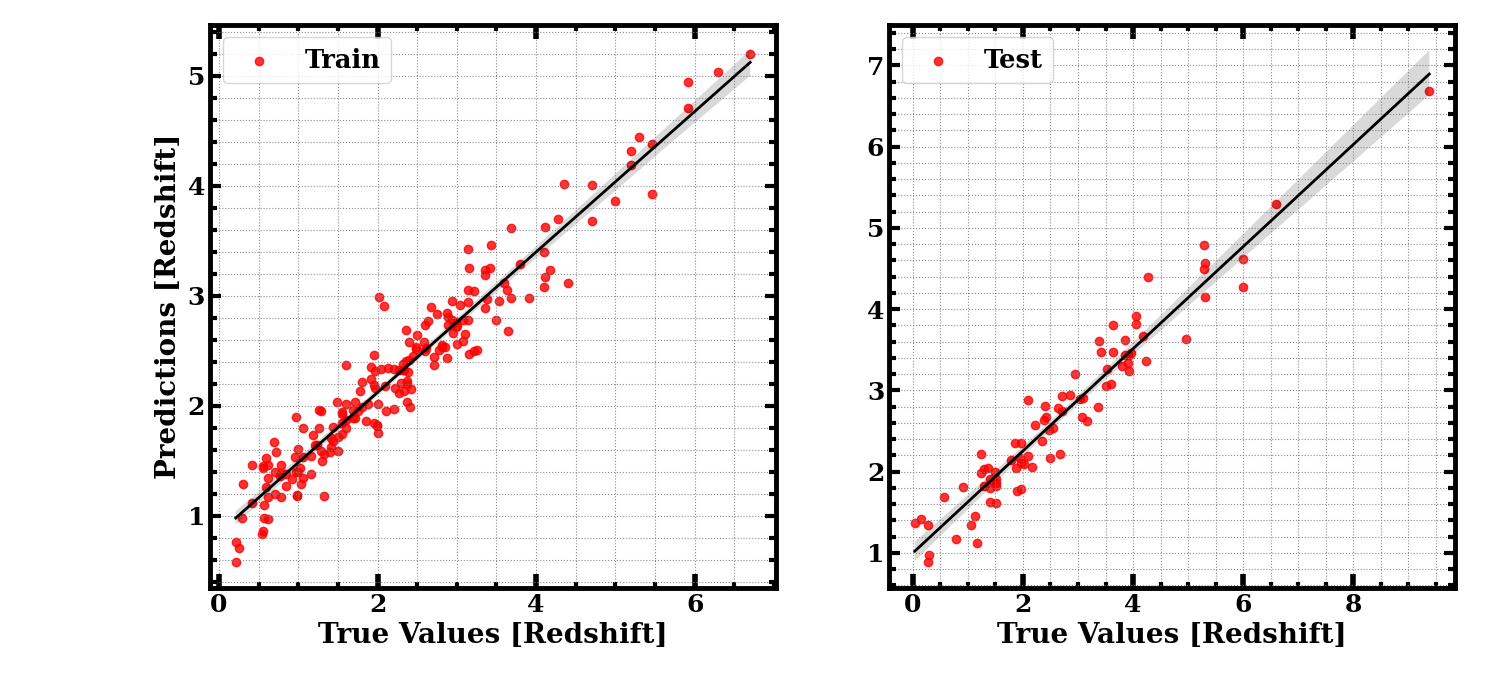}
\includegraphics[width=10.5cm]{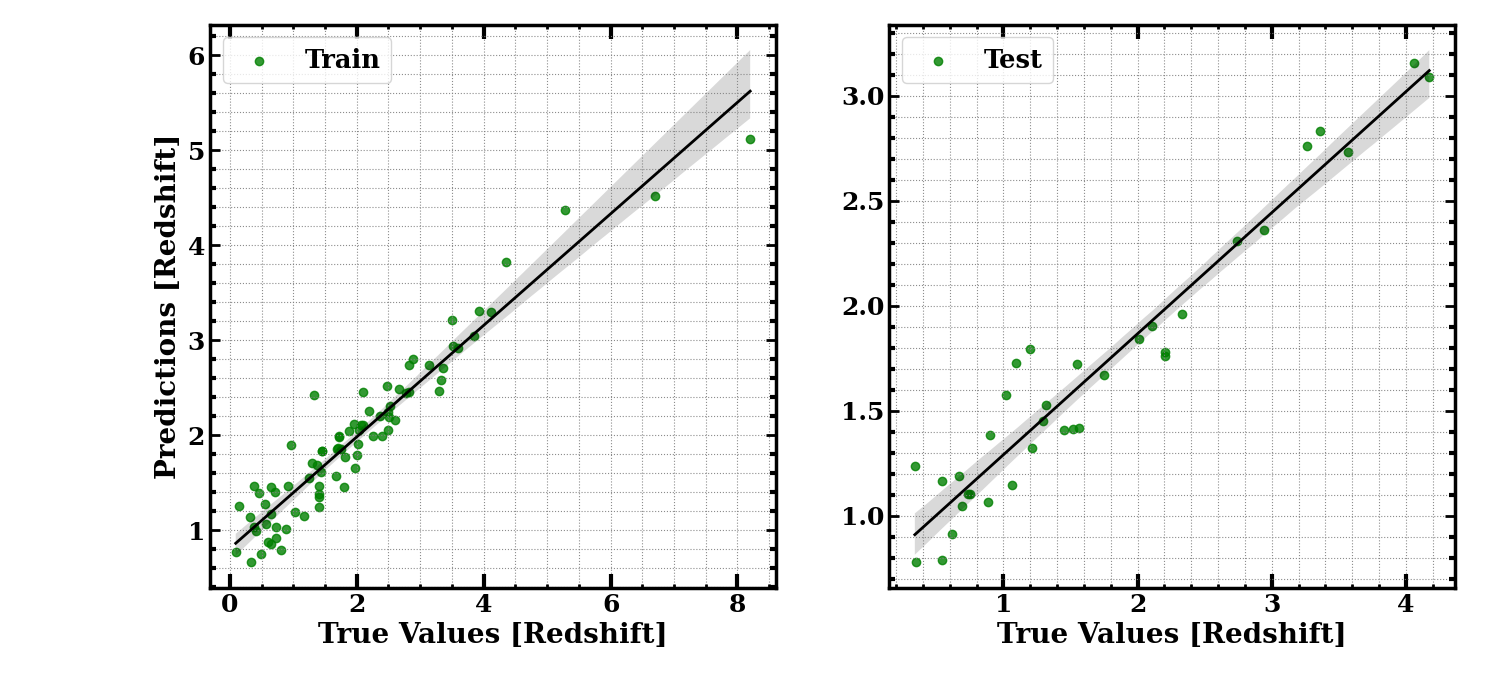}
\caption{The same as Fig.~\ref{fig:my_figure} but for the KW-only (top panels) and GBM-only (bottom panels) data sets for the ``Comp fluence and flux'' model.}
\label{fig:my_figure_2}
\end{center}  
\end{figure*}

We present the results of the GRB data fitting process using Ensemble Stacking method, combining DNNs and Random Forest (called {\it ensemble model} hereafter) in Table~\ref{tab:example_table}. The $r$-score coefficient ${R}^{2}$ and MAE are used as metrics to determine the goodness of the fit, with higher $R^2$ and lower MAE being preferred. The fit variance is shown separately for the train and test data sets for the KW and GBM, and for the combined KW-GBM data set, as described in Table~\ref{tab:example_table_0}. We consider both the Band and Compton spectral models for fluence and flux individually and in combination (fluence and flux). The best fit to test data was achieved using the ``Comp fluence and flux'' models when analyzing the KW and KW-GBM data, while the ``Comp flux'' model provided the best fit for the KW data. Note, however, that good fits were also achieved for the ``Comp flux'' model for the GBM and KW datasets. The results of the estimated pseudo redshift for these {\it ensemble model} are shown in Figure~\ref{fig:my_figure} for the best-fit and other models for the KW-GBM data sets. Figure~\ref{fig:my_figure_2} represents the best-fit models for the KW-only (top panels) and GBM-only (bottom panels) data sets. We have included the regression line with a sigma error, representing a 95\% confidence interval, in all the plots of the train and test data sets. The default error estimate is calculated using bootstrapped confidence intervals \citep{2008PaReL..29.1317H}. 

\begin{table}
    \caption{The $R^2$ coefficients and the mean absolute error (MAE) values for the {\it ensemble models, which combine DNNs and Random Forest through Ensemble Stacking method} have been calculated for each GRB spectral model using both the train and test data sets from KW, GBM, and KW-GBM. The bold text indicates the spectral model with the largest $R^2$ values in each GRB sample.}
	\centering
	\label{tab:example_table}
	\begin{tabular}{lcc} 
  \hline
	GBM data &  Train & Test \\
          & $ {R}^{2}$ -  MAE  & $ {R}^{2}$ -  MAE  \\
            \hline
		\hline
		Band fluence & 0.812 -  0.383 & 0.724 - 0.555 \\
        Band flux &  0.819 - 0.450 & 0.798 - 0.290 \\
        Comp fluence &  0.827 - 0.439 & 0.801 - 0.392 \\
         Comp flux & 0.823 - 0.384  & 0.812 - 0.437 \\
        Band fluence and flux & 0.844 - 0.370  & 0.821 - 0.498 \\
        Comp fluence and flux &  0.831 - 0.424  &   \bf0.823 - 0.343 \\
		\hline
		   KW-GBM data  &  Train & Test \\
     & $ {R}^{2}$ -  MAE  & $ {R}^{2}$ -  MAE \\
		\hline
		  Band fluence & 0.858 -  0.431 & 0.838 - 0.433 \\
        Band flux &  0.846 - 0.430 & 0.839- 0.413 \\
        Comp fluence &  0.857 - 0.409 & 0.836 - 0.433 \\
        Comp flux & 0.860 - 0.337 & 0.852 - 0.364 \\
        Band fluence and flux & 0.851 - 0.418  & 0.838 - 0.406 \\
        Comp fluence and flux &  0.861 - 0.369  &  \bf0.860 - 0.397 \\
		\hline
		   KW data &  Train & Test \\
     & $ {R}^{2}$ -  MAE  & $ {R}^{2}$ -  MAE   \\
		\hline
		Band fluence & 0.838 -  0.543 & 0.804 - 0.615\\
        Band flux &  0.827 - 0.503 & 0.766 - 0.457 \\
        Comp fluence &  0.826 - 0.480 & 0.810 - 0.509\\
        Comp flux &  0.842 - 0.441  &  \bf 0.831 - 0.361 \\
        Band fluence and flux & 0.846 - 0.463  & 0.812 - 0.551\\
        Comp fluence and flux &   0.838 - 0.406  &   0.829 - 0.512\\
		\hline
	\end{tabular}
\end{table}

%
%

\subsection{Pseudo-redshift estimates for {\it Fermi}-GBM data}
Based on the results in Table~\ref{tab:example_table}, we have applied the best-fit {\it ensemble models} to the {\it Fermi}-GBM data without known redshift to obtain the pseudo-redshift of those GRBs. Note that spectral information is not available for the KW GRBs with unknown redshifts. To evaluate how different samples of pseudo-redshifts compare with the samples of GRBs with true redshifts, we have performed two-sided Kolmogorov-Smirnov (KS) tests  \citep{1958ArM.....3..469H} between the samples. The resulting p-values are listed in Table \ref{tab:example_table2} against each model. 
Note that the highest p-value from the KS test does not coincide with the largest $R^2$ value of the ML model in Table~\ref{tab:example_table}. The difference between the corresponding $R^2$ values is not significant, however.
A p-value below 0.05 indicates that the pseudo-redshift sample is different from the true redshift sample at 95\% CL. Since none of the p-values reach this threshold in our analysis, we cannot reject the hypothesis that all of the pseudo-redshift samples are drawn from the same distributions as their respective true-redshift samples. A higher p-value indicates more resemblance between the true- and pseudo-redshift samples. For subsequent analyses we have thus chosen the pseudo-redshift samples for which the p-values are the largest.

\begin{figure}
  \centering
  \includegraphics[width=0.9\columnwidth]{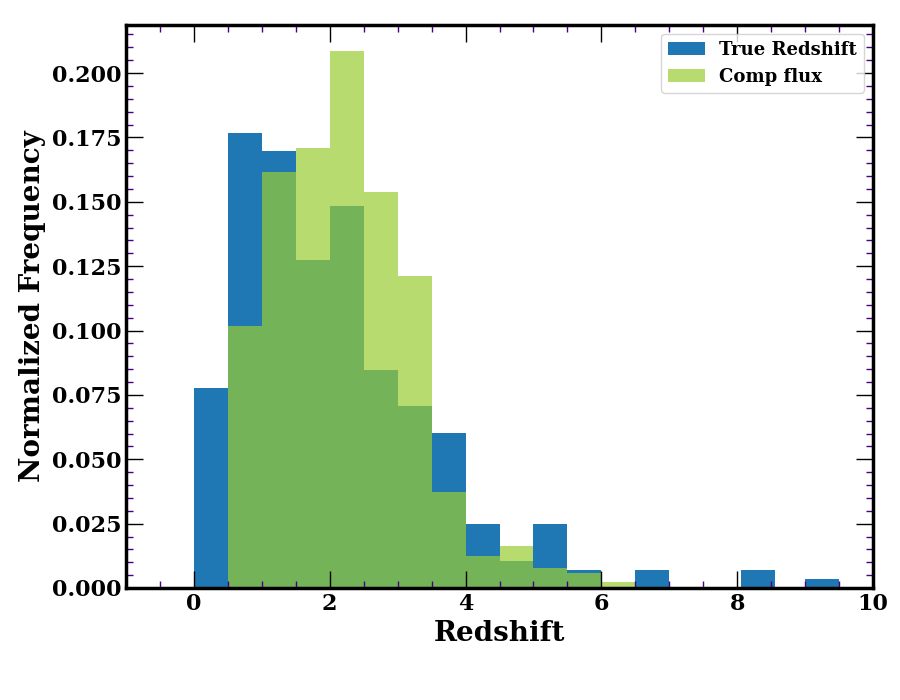}
  \includegraphics[width=0.9\columnwidth]{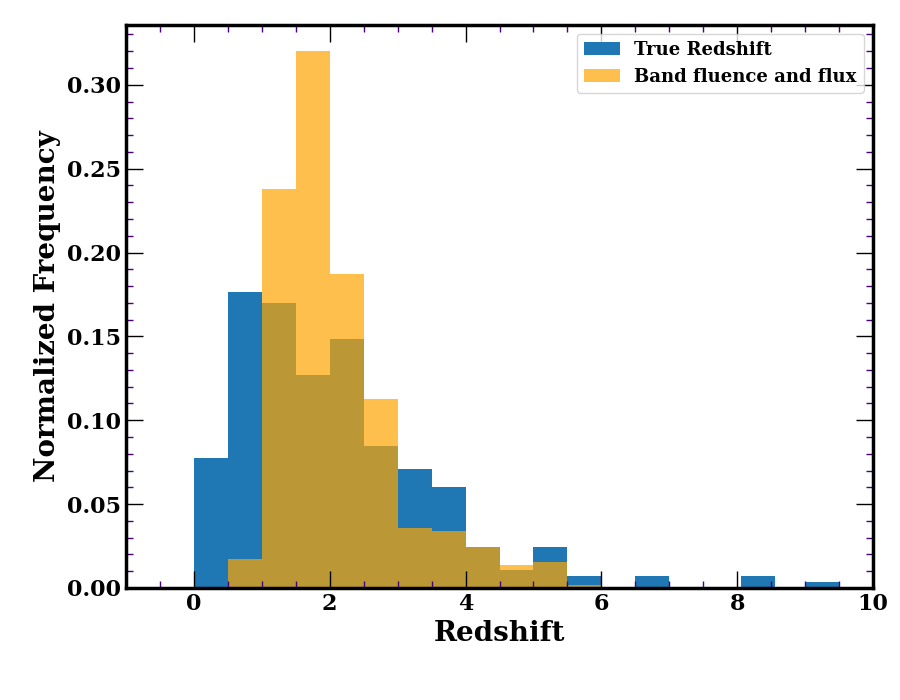}
   \caption{
   Normalized histograms comparing GRB samples with ``True-Redshift'' and pseudo-redshift obtained with different {\it ensemble models.} {\it Top panel --} Comparison between the ``True Redshift'' of 226 KW-GBM GRBs and pseudo-redshift of 1708 GBM GRBs obtained using the ``Comp flux'' model.  The p-value from the KS test for this pseudo-redshift sample is 0.8319. 
   {\it Bottom panel --} Comparison between the ``True Redshift'' of 183 KW-GBM GRBs and pseudo-redshift of 647 GBM GRBs obtained using the ``Band fluence and flux'' model.  The p-value from the KS test for this pseudo-redshift sample is 0.5713. See Table~\ref{tab:example_table} for p-values.}
  \label{fig:my_figure_3}
\end{figure}

We show normalized distributions of pseudo-redshift and true-redshift for comparisons in Fig.~\ref{fig:my_figure_3}, where the pseudo-redshift samples were derived from the ``Comp flux'' model (top panel) and  ``Band fluence and flux'' model (bottom panel); both applied to the KW-GBM data sets. These are the two best-fit DNN models for pseudo-redshift estimation as the p-value from the KS-test for these samples are the highest, see Table~\ref{tab:example_table} for details. Note that the number of GRBs in these samples varies because not all spectral information is available for each GRB. See Table~\ref{tab:example_table_0} for details of the number of GRBs in each sample for both true-redshift and pseudo-redshift.

\begin{table}
    \caption{p-values from a two-sided Kolmogorov-Smirnov (KS) test between the GRB samples with measured redshift and estimated pseudo redshift for GBM data without measured redshift. The ML models for estimating pseudo redshift is the same as in Table~\ref{tab:example_table}.}
	\centering
	\label{tab:example_table2}
	\begin{tabular}{lc} 
  \hline
		   GBM data & p-value \\
            \hline
		\hline
	Band fluence & 0.1532\\
        Band flux & 0.0681\\
        Comp fluence & 0.1531\\
         Comp flux & \bf 0.5713\\
        Band fluence and flux & 0.1532\\
        Comp fluence and flux & 0.1531\\
		\hline
		   KW-GBM data & p-value\\
		\hline
		  Band fluence & 0.0681\\
        Band flux & 0.0681\\
        Comp fluence & 0.0948\\
        Comp flux & \bf 0.8319\\
        Band fluence and flux & 0.5713\\
        Comp fluence and flux &  0.1745\\
		\hline
	\end{tabular}
\end{table}

\subsection{Phenomenological correlations with the true-redshift and pseudo-redshift samples}
\label{sec:section5}
After generating pseudo-redshift GRB samples using the two best-fit {\it ensemble models}, namely the ``Compton flux'' and ``Band fluence and flux'', we proceed to explore the Amati \citep{Amati2002_81A, Amati2006_372} and Yonetoku \citep{2004ApJ...609..935Y, 2014Ap&SS.351..267Z} correlations. Here we follow the same procedure outlined by \citet{Dirirsa2019, 2017heas.confE...2D}. 
The choice to include only the Amati and Yonetoku relations in our study was based on availability of relevant data. We used the {\it Fermi}-GBM and Konus-{\it Wind} catalogs for our study and these catalogs include enough information for the Amati and Yonetoku relations. Other relations, e.g., the Ghirlanda relation \citep{Ghirlanda2004_616} requires information on the GRB jet opening angle, or the spectral-lag luminosity relation \citep{Norris2000ApJ...534..248N} that requires information on difference in time of arrival of high- and low-energy photons. Those information are not available in these catalogs.

We calculate the isotropic-equivalent gamma-ray energy release $E_{\rm iso}$ using the measured bolometric fluence ${S}_{\rm bolo}$ as
\begin{eqnarray}\label{Eq:Eiso}
E_{\rm iso} = \frac{4\pi d_{\rm L}^2}{1+z} {S}_{\rm bolo},
\end{eqnarray}
where ${S}_{\rm bolo}$ is calculated using either the Band model in equation~(\ref{eq:Band}) or the Comptonized model in equation~(\ref{eq:Comp}) for the spectrum and reported in the GBM and KW catalogs. The intrinsic peak energy of the ${\rm \nu}f_{\rm \nu}$ spectrum is $E_{\rm i,p} = E_{\rm p}(1+z)$. We calculate the luminosity distance $d_L$ in equation~(\ref{Eq:Eiso}) using a flat $\Lambda$CDM cosmological model as
\begin{eqnarray}\label{Eq:dl}
d_{\rm L} = (1+z)\frac{c}{H_{\rm 0}}\int_0^z \frac{dz'}{\sqrt{(1-\Omega_{\Lambda})(1+z')^3+\Omega_{\Lambda}}}.
\end{eqnarray} 
where $H_0 = 67.3 \, \text{km s}^{-1} \, \text{Mpc}^{-1}$ and $\Omega_{\Lambda} = 0.685$ \citep{Planck_Collaboration2018}. 
We calculate the isotropic-equivalent peak luminosity $L_{\rm iso}$ using the measured peak bolometric flux $P_{\rm bolo}$ as
\begin{eqnarray}\label{Eq:Liso}
L_{\rm iso} = 4\pi d_{\rm L}^2\ {P}_{\rm bolo},
\end{eqnarray} 
Where ${P}_{\rm bolo}$ is calculated using either the Band or Comptonized model for the spectrum during the peak flux and reported in the GBM and KW catalogs. Note that $E_{\rm i,p}$, in this case, is calculated from the peak of the ${\rm \nu}f_{\rm \nu}$ spectrum during the peak flux. 

The Amati and Yonetoku correlations are given by  
\begin{eqnarray}\label{Eq10}
E_{\rm iso} = k \left(\displaystyle\frac{E_{\rm i,p}}{E_{\rm 0}} \right)^m \,\,\, {\rm and} \,\,\,\, L_{\rm iso} = k \left(\displaystyle\frac{E_{\rm i,p}}{E_{\rm 0}} \right)^m,
\end{eqnarray}
respectively, where $E_{\rm 0}$ is a reference energy. Parameters $m$ and $k$ are the index and intercept, respectively. We compute $E_{\rm iso} - E_{\rm i,p}$ and $L_{\rm iso} - E_{\rm i,p}$ combinations for each GRB in our samples using the standard cosmological parameters described above.     
We use a linearized form of these correlations $y = mx + k$ for fitting data, where
\begin{eqnarray}\label{Eq11}
x \equiv \log_{10} \left( \frac{E_{\rm i,p}}{E_{0}} \right) \,,\,
y \equiv \log_{10} \left( \frac{E_{\rm iso}}{\rm erg/s} \right)\, {\rm or}\, 
\log_{10} \left( \frac{L_{\rm iso}}{\rm erg/s} \right)\, .
\end{eqnarray}
We calculate the uncertainty on $y$ using the following formula \citep{Wang2016_585, Demianski2017_693}
\begin{eqnarray}\label{Eq12}
\sigma_{y} = \sqrt{\sigma_{k}^2+m^2\sigma_{x}^2+\sigma_{m}^2x^2+\sigma_{\rm ext}^2}\,,
\end{eqnarray}
where $\sigma_{\rm ext}$ is extrinsic uncertainty on $y$, which is treated as an unknown parameter. The error $\sigma_{x}$ is calculated from the reported error on $E_{\rm p}$ in the catalogs. To determine the parameters \(k\), \(m\), and \(\sigma_{\text{ext}}\) 
we maximize the likelihood function \citep{2005physics..11182D}
\begin{eqnarray}\label{Eq13}
\begin{aligned}
{L}(m,k,\sigma_{\rm ext}) = {} & \displaystyle\frac{1}{2}\sum_{{i}}^N \ln (\sigma^2_{\rm ext}+\sigma^2_{y_{\rm i}}+m^2\sigma^2_{x_{\rm i}}) 
 \\
 &  + \displaystyle\frac{1}{2} \sum_{{i}}^N{\displaystyle\frac{(y_{\rm i}-mx_{\rm i}-k)^2}{(\sigma^2_{\rm ext}+\sigma^2_{\rm y_{i}}+m^2\sigma^2_{x_{\rm i}})}}.
\end{aligned}
\end{eqnarray}
We compute errors on the fitting parameters $k$, $m$ and $\sigma_{\rm ext}$ following \citet{2005physics..11182D}.

\begin{figure}
\begin{center}
\includegraphics[width=0.9\columnwidth]{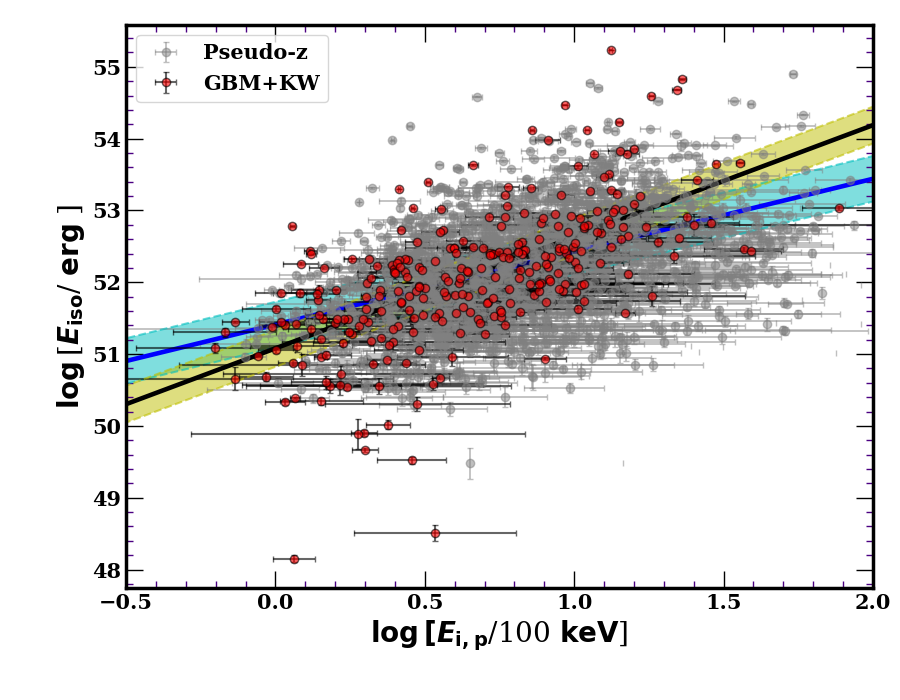}
\includegraphics[width=0.9\columnwidth]{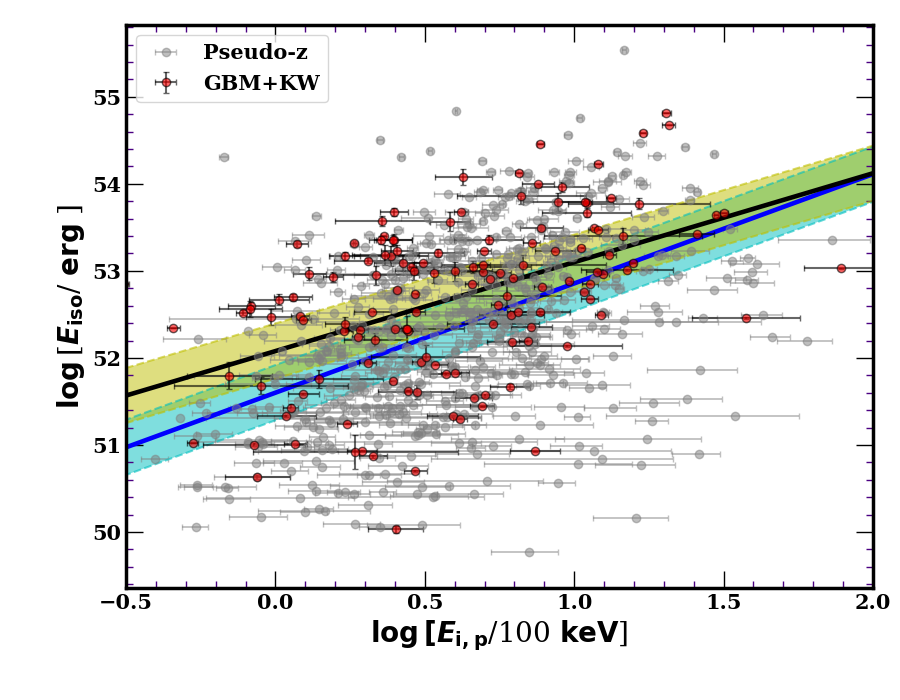}
\caption{The Amati relation fits to the GRB samples with true redshift (red-filled circles) and pseudo-redshift (gray-filled circles).  The {\it ensemble models} used for predicting pseudo redshift is ``Comp flux'' derived from KW-GBM true-redshift data (see Table~\ref{tab:example_table}, KS-test p-value: 0.839). The black and blue solid lines correspond to the fits to the KW-GBM true-$z$ sample and GBM pseudo-$z$ sample, respectively. The shaded region in each case represents 90\% confidence level to the fit. The fit parameters are listed in Table~\ref{tab:4}. {\it Top panel --} $E_{\rm iso}$ calculated using Compton fluence (1708 GRBs with pseudo-$z$). {\it Bottom panel --} $E_{\rm iso}$ calculated using Band fluence (621 GRBs with pseudo-$z$).
} 
\label{fig:my_figure_9}  
\end{center}
\end{figure}
%
\begin{figure}
\begin{center}
\includegraphics[width=0.9\columnwidth]{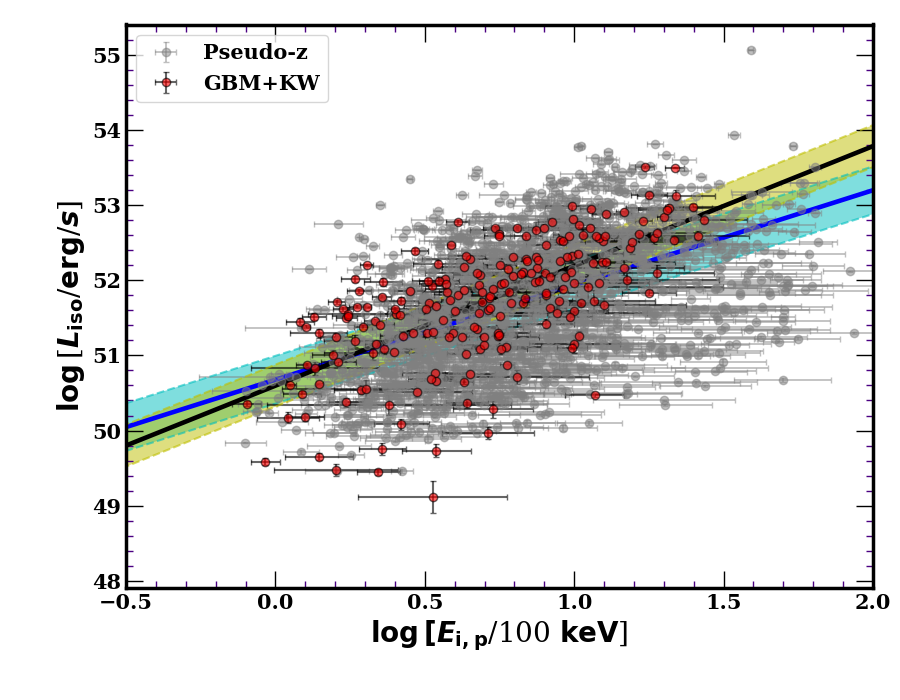}
\includegraphics[width=0.9\columnwidth]{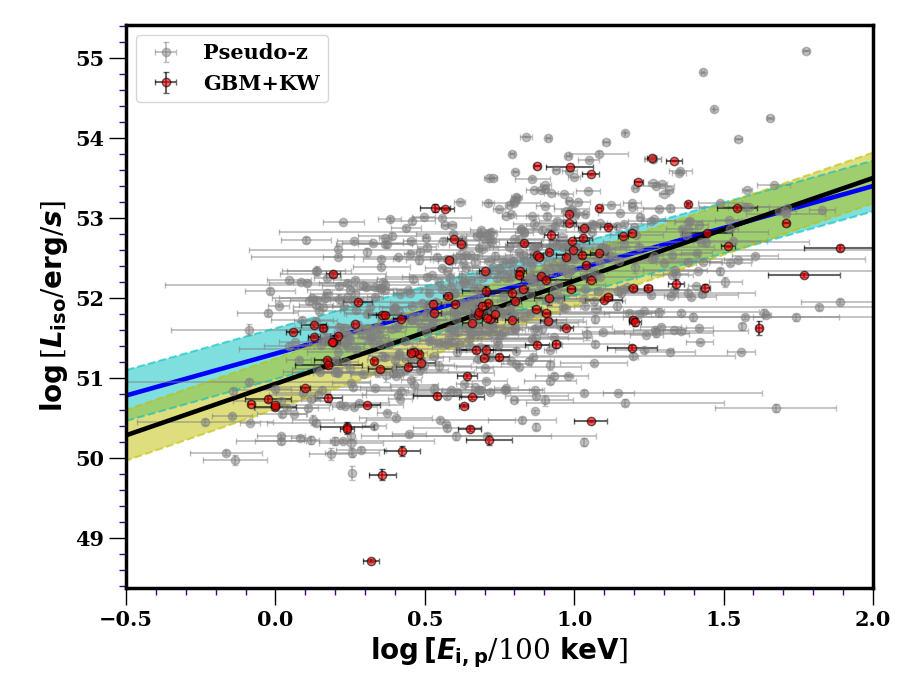}
\caption{The Yonetoku relation fits the GRB samples with true redshift (red-filled circles) and pseudo-redshift (gray-filled circles). The {\it ensemble model} used for predicting pseudo redshift is ``Comp flux'' derived from KW-GBM true-redshift data (see Table~\ref{tab:example_table}, KS-test p-value: 0.839). The black and blue solid lines correspond to the fits to the KW-GBM true-$z$ sample and GBM pseudo-$z$ sample, respectively. The shaded region in each case represents 90\% confidence level to the fit. The fit parameters are listed in Table~\ref{tab:4}. {\it Top panel --} $L_{\rm iso}$ calculated using Compton flux (1708 GRBs with pseudo-$z$). {\it Bottom panel --} $L_{\rm iso}$ calculated using Band flux (621 GRBs with pseudo-$z$).}
\label{fig:my_figure_10}  
\end{center}
\end{figure}
In Figs.~\ref{fig:my_figure_9} and \ref{fig:my_figure_10}, we have plotted the Amati and Yonetoku relations, respectively, for the GRB samples with true redshift and pseudo redshift. We have calculated ${E}_{\rm iso}$ and ${L}_{\rm iso}$ for the KW-GBM samples with known redshift and the GBM sample (with unknown redshift) using pseudo redshift from the ``Comp flux''  model, which gave a p-value of 0.8319 for a KS test (see Table~\ref{tab:example_table}). We found that the ``Band fluence and flux''  model for pseudo-redshift gives a large scatter in ${E}_{\rm iso}$ and ${L}_{\rm iso}$ values. The errors on ${E}_{\rm iso}$, ${L}_{\rm iso}$ and $E_{\rm i,p}$ plotted in the figures correspond to the reported errors on fluence, flux and the peak photon energy, respectively. We have also plotted the Amati relation fit and the Yonetoku relation fit in Figs.~\ref{fig:my_figure_9} and \ref{fig:my_figure_10}, respectively, along with 68\% confidence level error on the fit (shaded area). In Table~\ref{tab:4} we list the fit parameters along with their $1\sigma$ errors as well as the decorrelation energy $E_{0,\rm dec}$, at which there is no correlation between the parameters $k$ and $m$ \citep[see, e.g.,][]{Dirirsa2019}. We have replaced $E_0 \to E_{0,\rm dec}$ in equation~(\ref{Eq11}). Note that the error on the fit $\sigma_y$ plotted in the figures is dominated by the extrinsic error $\sigma_{\rm ext}$, as also found by \citet{Dirirsa2019}. 

\begin{table*}
\centering
\caption{Results of the Amati \citep{Amati2002_81A} and Yonetoku \citep{2004ApJ...609..935Y} correlation fits applied to the KW-GBM samples of GRBs with true redshift and GBM sample of GRBs with pseudo redshift. In all cases, pseudo redshift was found from the ML model using ``Comp flux''  spectral model.}
\begin{tabular}{c|ccccc}
\hline
       \bf Amati correlation & Samples & ${E}_{0, \rm dec}$ (keV) & ${k}$& ${m}$ & $\sigma_{\rm ext}$\\
         \hline
        \hline
         Band fluence (true $z$) &  GBM+KW & 977 & 52.08 $\pm$ 0.12 & 1.02 $\pm$ 0.16 & 0.80 $\pm$ 0.05\\
         Band fluence (pseudo $z$) &  GBM & 973 & 51.60 $\pm$ 0.07 & 1.30 $\pm$ 0.10 & 0.90 $\pm$ 0.03\\ 
         Comp fluence (true $z$) &  GBM+KW & 103 & 51.08 $\pm$ 0.08 & 1.55 $\pm$ 0.11 & 0.72 $\pm$ 0.03\\
         Comp fluence (pseudo $z$) &  GBM & 100 & 51.41 $\pm$ 0.10 & 1.02 $\pm$ 0.12 & 0.66 $\pm$ 0.01\\
         \hline
         \bf Yonetuko correlation &  Samples & ${E}_{0, \rm dec}$ (keV) & ${k}$& ${m}$ & $\sigma_{\rm ext}$ \\
         \hline
         Band flux (true $z$) &  GBM+KW & 104 & 50.93 $\pm$ 0.14 & 1.30 $\pm$ 0.16 & 0.70 $\pm$ 0.05\\
         Band flux (pseudo $z$) &  GBM & 973 & 51.30 $\pm$ 0.10 & 1.05 $\pm$ 0.08 & 0.70 $\pm$ 0.02\\
         Comp-flux (true $z$) &  GBM+KW & 964 & 50.60 $\pm$ 0.10 & 1.60 $\pm$ 0.12 & 0.60 $\pm$ 0.03\\
         Comp-flux (pseudo $z$)  & GBM & 100 & 50.70 $\pm$ 0.02 & 1.30 $\pm$ 0.02 & 0.73 $\pm$ 0.02\\
         \hline 
\end{tabular}
         \label{tab:4}
\end{table*}

We found that the value of the Amati and Yonetoku correlation parameters vary depending on the spectral model even for the GRB samples with true redshift. The values obtained for the pseudo-redshift sample also vary from those obtained for the true-redshift samples. The ranges of the Amati correlation fit parameter values we have obtained are $k = 51.0 - 52.2$ and $m = 0.90-1.66$. As for comparisons of our results with previous works (all using true-redshift GRB samples only), \citet{Dirirsa2019} found $k = 53.31 \pm 0.07$ and $m = 1.15 \pm 0.16$ for their F10+W2016 sample and \citet{Demianski2017_693} found $k = 52.53 \pm 0.02$ and $m = 1.75 \pm 0.18$. Although these values slightly differ from each other, the main uncertainty in all the fits arises from a somewhat large extrinsic error $\sigma_{\rm ext}$. In the case of the Yonetoku correlation fit parameter values, the ranges are $k = 50.5 - 51.4$ and $m = 0.97-1.72$. Again, while comparing with previous work,  \citet{2017heas.confE...2D} found $k = 51.35 \pm 0.24$ and $m = 1.88 \pm 0.17$. For the Yonetoku correlation as well, $\sigma_{\rm ext}$ dominates. 

\begin{figure}
\begin{center}
\includegraphics[width=\columnwidth]{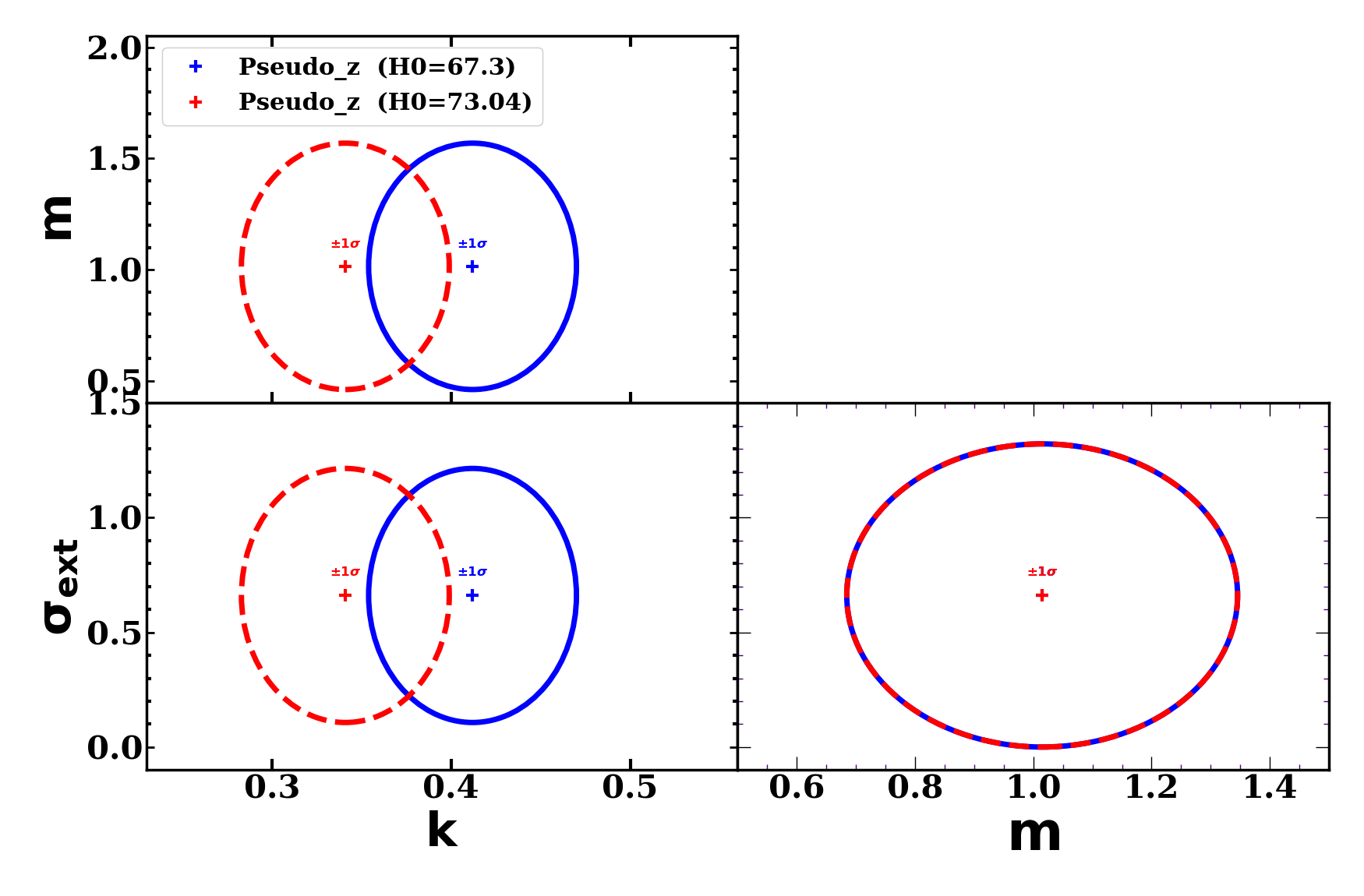}
\includegraphics[width=\columnwidth]{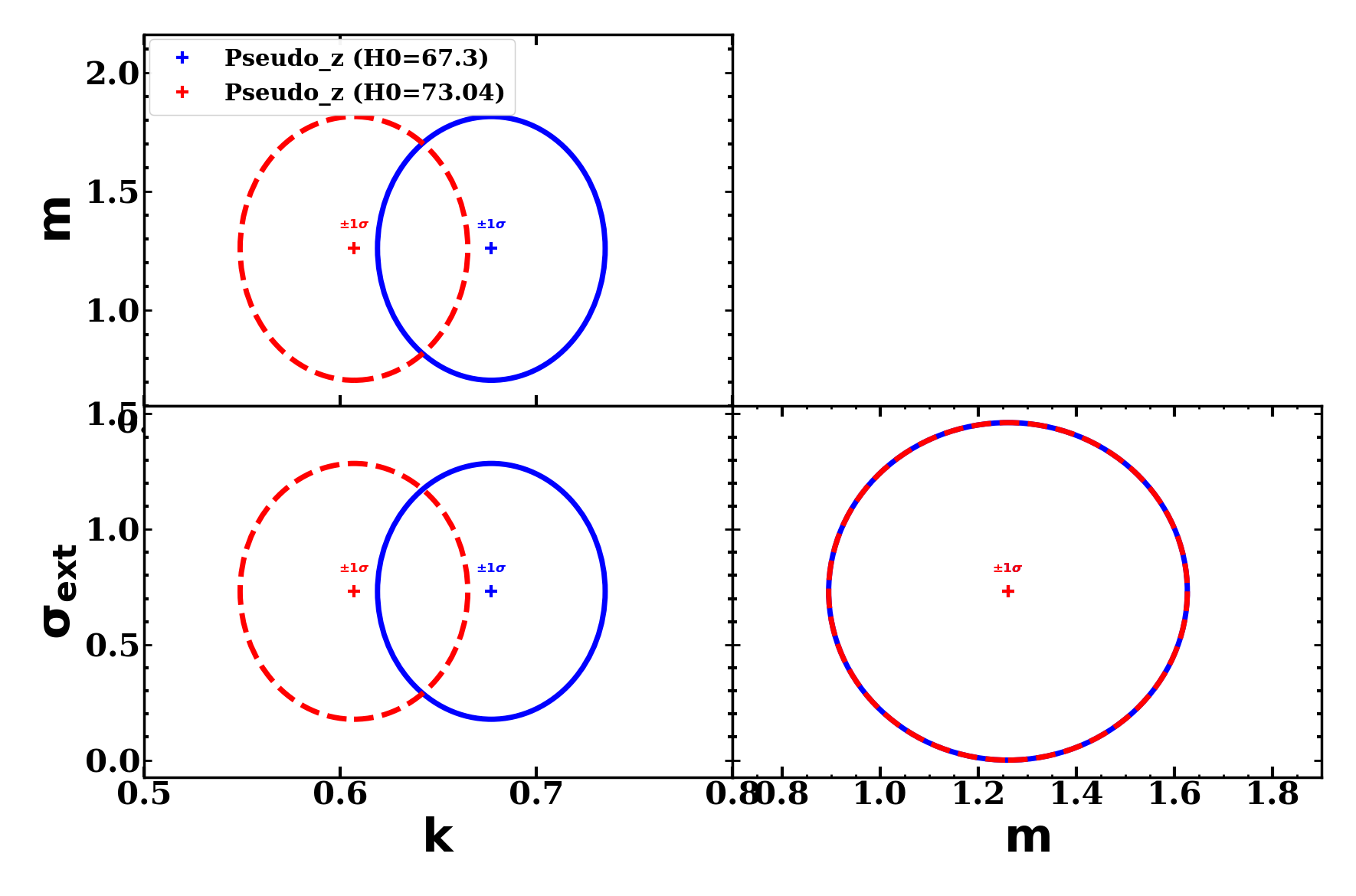}
\caption{{The best-fit parameters, with their $1\sigma$ errors, of the Amati relation (top panel) and Yonetoku relation (bottom panel) for the pseudo-redshift sample of GRBs, obtained using the ``Comp flux'' model derived from KW-GBM true-redshift data. The blue solid curve is for $H_0 = 67.3$~km~s$^{-1}$ Mpc$^{-1}$ from Plank data \citep{Planck_Collaboration2018} and the red dashed curves are for $H_0 = 73.04$~km~s$^{-1}$ Mpc$^{-1}$ from the baseline fit by \citet{2022ApJ...934L...7R}.}}
\label{fig:my_figure_11}  
\end{center}
\end{figure}

Note that the values of the Amati and Yonetoku relation parameters listed in Table~\ref{tab:4} also depend on the cosmological parameters $H_0$ and $\Omega_\Lambda$ \citep[see, e.g.,][]{Dirirsa2019}. As an illustration and in view of the ``Hubble tension'', we have fitted the Amati and Yonetoku correlations using $H_0 = 73.04$~km~s$^{-1}$ Mpc$^{-1}$ according to the baseline model by \citet{2022ApJ...934L...7R}. The changes in the fit parameters are shown in Figure \ref{fig:my_figure_11}.

\section{Summary and Discussion} \label{sec:discussion}

Our analysis showcases the results of GRB pseudo redshift estimation using the regression technique in deep neural networks. We have used the {\it Fermi}-GBM and Konus-Wind catalogs with known redshift to train and test our {\it ensemble models}. We have used only observable quantities such as the GRB duration and gamma-ray spectral fit parameters as inputs (see Table~\ref{tab:table one} for different input parameters and Table~\ref{tab:example_table_0} for different GRB samples). The obtained results demonstrate that different spectral fits yield varying degrees of fit with the {\it ensemble models} (see Figs.~\ref{fig:my_figure} and \ref{fig:my_figure_2}). The goodness of our results is evaluated based on the coefficient of determination ($R^2$) and the minimization of the mean absolute error (MAE) in the linear regression models algorithm. Table~\ref{tab:example_table} provides a comprehensive breakdown of the results achieved, presenting the $R^2$ and MAE values for each model. Notably, the high coefficient values observed in the test set are attributed to the {\it ensemble models} using Compton spectral fits to the flux and fluence data. It is worth mentioning that our analysis is based on a limited number of available predictors, specifically 126 GBM LGRBs and 338 KW LGRBs with known redshift. 

It is, however, well known that ML techniques have significant drawbacks to uncover explicit mathematical relations between different input parameters. Especially in the case of DNNs, there is a lack of integrability; the mathematical relationship among the input parameters is not understood. Notably, other ML algorithms such as Linear Regression and Gaussian Process Regression exhibit better integrability than DNNs. Despite this, we observed that those other models fail to provide a satisfactory fit to the data that we have analyzed.

We compared pseudo-redshift samples of GBM data obtained by applying {\it ensemble models} with true-redshift samples using two-sided Kolmogorov-Smirnov tests. The resulting p-values are listed in Table~\ref{tab:example_table}. Since none of the p-values is $<0.05$, the null hypothesis that the corresponding pseudo-redshift and true-redshift samples are from the same distribution can be rejected at 95\% CL. The largest p-values obtained are for the ``Comp flux'' and ``Band fluence and flux'' models with corresponding histograms shown in Fig.~\ref{fig:my_figure_3}. Subsequently, we focus on the pseudo-redshift sample of 1708 GBM-GRBs obtained with the ``Comp flux''  model.    
Within the analyzed data, the `Comp flux''  model achieves good redshift estimations within the range of $0.5 \lesssim z \lesssim 6$. This range is determined based on the distribution of true redshift values, which peaks between $z \approx 0.5$ and $2.5$. Future detection of more LGRBs, especially in the low- and high-redshift ranges will further improve our estimate of pseudo redshift.

Next, we have used 1708 GRBs with pseudo redshift to verify the Amati \citep{Amati2002_81A} and Yonetoku \citep{2004ApJ...609..935Y} phenomenological correlations using a likelihood fit procedure. We found that the pseudo-redshift sample does satisfy these phenomenological correlations as do the true-redshift samples (see Figs.~\ref{fig:my_figure_9} and \ref{fig:my_figure_10}), although both have large scatter due to unknown physical origin of these correlations embedded in the parameter $\sigma_{\rm ext}$ (see Table \ref{tab:4}). A similar conclusion was drawn by previous studies \citep{Wang2016_585, Demianski2017_693, Dirirsa2019}. The normalization ($k$) and slope ($m$) parameters vary somewhat between the true- and pseudo-redshift samples and between the bolometric flux or fluence values used in the analysis (either Band or Compton fit to the spectra) for the same true- or pseudo-redshift sample. In general, the pseudo-redshift sample reasonably reproduces the Amati and Yonetoku correlations.  

\section{Conclusions} \label{sec:conclusions}
Our study introduces a novel approach using deep-learning techniques and regression tools to estimate the redshift of GRBs based on data from the {\it Fermi}-GBM and Kouns-{\it Wind} instruments. By employing a deep-learning algorithm, we aim to overcome the limitations of traditional regression methods and enhance the accuracy of our predictions. In our approach, we adopt the stack method, a powerful ensemble technique that combines the outcomes of multiple DNN models and a Random Forest meta-model. This strategy is employed to mitigate overfitting and enhance the generalization capabilities of our analysis. By combining the strengths of different models, we aim to achieve more robust and reliable predictions of redshifts for GRBs, ensuring our analysis remains accurate and avoids biases from initial results

Initial findings from our deep neural network models are promising, as the estimated redshift distribution follows the true redshift distribution. These preliminary results showcase the potential of our approach in accurately estimating the redshifts of GRBs. The success of the DNN models indicates that they capture the underlying patterns and relationships in the data, offering a promising avenue for further exploration and refinement. By obtaining a large number of pseudo-redshifts, we were able to test phenomenological correlations involving cosmological model-dependent quantities such as the isotropic energy release and peak isotropic luminosity. Therefore the pseudo-redshift sample we have obtained holds promise for using GRBs as cosmological standard candles.

\section*{Acknowledgements}\label{sec:section7}
We thank Feraol Fana Dirirsa, Judith Racusin, Oleg Kargaltsev, Dan Kocevski, and Natalie Nuessle for the helpful discussion. We acknowledge support from the Organization for Women in Science for the Developing World (OWSD); the National Research Foundation (NRF), South Africa,  BRICS STI programme; the National Institute for Theoretical Computational Sciences (NITheCS), South Africa; and the South African Gamma-ray Astronomy Programme (SA-GAMMA).
\section*{Data Availability}
\label{sec:section8}
The {\it Fermi}-GBM data used in this study can be accessed through the official website of the {\it Fermi} Gamma-ray Burst Monitor (GBM) mission. The website provides a comprehensive archive of GBM data, including spectral information, light curves, and other relevant data products. Similarly, the Kouns-Wind Gamma-ray Burst (KW-GRB) data used in this study is available through the dedicated KW-GRB data repository. The repository offers a collection of GRB data observed by the Kouns-Wind instrument, providing researchers access to a rich dataset for further analysis and investigation.



\bibliographystyle{mnras}
\bibliography{example} 




\appendix

\section{}
\label{app:DNN}

We describe here details of different elements of a DNN used in our analyses.

\begin{enumerate}

\item {\bf Network layers:} 
In DNNs, each layer, including hidden layers (mostly non-linear) described by ${h}^{(I)} (x)$ with predominantly non-linear operations \citep{LIM20221825}, contributes to the overall network functionality. Here ${h}$ refers to the activation or output of a specific layer in the network. It represents the transformed information that is propagated through the network. On the other hand, ${I}$ denotes the layer index or number, indicating the position of the layer within the network architecture. Lastly, ${x}$ represents the input to the DNN, which can be a single data point or a batch of data points serving as the starting point for the network's computation.

\item {\bf Activation Function:} Activation functions play a crucial role in the training process and overall effectiveness of NNs by adjusting the outputs. In the case of DNNs, non-linear activation functions are typically used to facilitate the creation of complex non-linear mappings between the input and output data. Some commonly used activation functions include the Sigmoid, Tanh, Softmax, and Rectified Linear Unit (ReLU) \citep{2017arXiv171011272A}.

In this analysis, we employ the ReLU activation function \citep{geron2022hands}, which is widely used in neural networks due to its effectiveness in constructing efficient DNNs \citep{2022arXiv220411786C}. It is advantageous as it allows for efficient training and computation in DNNs. It helps in addressing the vanishing gradient problem by avoiding saturation of gradients and enabling the network to learn non-linear relationships effectively. The ReLU activation function is defined as:
\begin{eqnarray}
\label{equ3}
 Relu(z) = max(0,z)
\end{eqnarray}
\begin{eqnarray}
\label{equ4}
f(x) = (0,{w}_{0} + {w}_{1} + {w}_{2} + ... + {w}_{m} + b) 
\end{eqnarray} 
\begin{eqnarray}
\label{equ5}
{h}_{w,b} (X) = max(X.w + b,0)
\end{eqnarray}
where $z$ in equation~(\ref{equ3}) represents the input to the activation function. In equation~(\ref{equ4}), ${w}$ and ${b}$ represent the weight and bias, respectively. These parameters are crucial for the NN model as they determine the strength of the connections between the neurons and the offset introduced at each neuron. The weight ${w}$ controls the influence of each input feature on the output prediction, while the bias ${b}$ adjusts the prediction by providing an additional constant term. These are initially assigned random values and then updated during the training process to optimize the NN model. This iterative adjustment enables the model to learn from the data and to make more accurate predictions over time. Equation~(\ref{equ5}) signifies that the layer's output is obtained by taking the dot product of the input $X$ with the $w$, adding the $b$, and by applying the ReLU activation function to ensure non-negativity.

\end{enumerate}

\bsp	
\label{lastpage}
\end{document}